\documentclass[iop]{emulateapj}
\usepackage{natbib}
\shorttitle{The Stellar Orbital Structure in Axisymmetric Galaxy Models with SMBHBs}
\shortauthors{Li Holley-Bockelmann and Khan}

\begin{document}

\title{The Stellar Orbital Structure in Axisymmetric Galaxy Models with Supermassive Black Hole Binaries}


\author{Baile Li\altaffilmark{1}, Kelly Holley-Bockelmann\altaffilmark{2,3}, and Fazeel Mahmood Khan\altaffilmark{4}}
\affil{
\altaffilmark{1}{Key Laboratory for Research in Galaxies and Cosmology, Shanghai Astronomical Observatory, Chinese Academy of Sciences, 80 Nandan Road, Shanghai
200030, China; lib@shao.ac.cn}\\   
\altaffilmark{2}{Department of Physics and Astronomy, Vanderbilt University,
    Nashville, TN 37235, USA; k.holley@vanderbilt.edu}\\     
\altaffilmark{3}{Department of Physics, Fisk University, Nashville, TN 37208, USA}\\
\altaffilmark{4}{Department of Space Science, Institute of Space Technology, P.O. Box 2750 Islamabad, Pakistan; khanfazeel.ist@gmail.com}
}

\begin{abstract}

It has been well-established that particular centrophilic orbital families in non-spherical galaxies can, in principle, drive a black hole binary to shrink its orbit through three-body scattering until the black holes are close enough to strongly emit gravitational waves. Most of these studies rely on orbital analysis of a static SMBH-embedded galaxy potential to support this view; it is not clear, however, how these orbits transform as the second SMBH enters the 
center, so our understanding of which orbits actually interact with a SMBH binary is not ironclad.
Here, we analyze two flattened galaxy models, one with a single SMBH and one with a binary, to determine 
which orbits actually do interact with the SMBH binary and how they compare with the set predicted in single SMBH-embedded models.
We find close correspondence between the centrophilic orbits predicted to interact with the binary and those that are actually scattered by the binary, in terms of energy and $L_{z}$ distribution, where $L_{z}$ is the z component of a stellar particle's angular momentum. Of minor note: because of the larger mass, the binary SMBH has a radius of influence $\sim 4$ times larger than in the single SMBH model, which allows the binary to draw from a larger reservoir of orbits to scatter. Of the prediction particles and scattered particles, nearly half have chaotic orbits, 40\% have fx:fy=1:1 orbits, 10\% have other resonant orbits.


\end{abstract}


\section{Introduction}
Supermassive black holes (SMBHs), with masses in the range of $10^{6} M_{\odot}$ to $10^{9} M_{\odot}$, are a part of nearly every galaxy center \citep{1995ARA&A..33..581K,2013ARA&A..51..511K}. As galaxies grow through merging, the SMBHs will sink to the center of the remnant, eventually forming a SMBH binary (SMBHBs) \citep{1980Natur.287..307B} that will coalesce -- though the timescale from galaxy merger to black hole merger could be more than a Hubble time. The process that governs SMBHB coalescence can be broken into three stages.  First, dynamical friction from the background stars and gas drives the two black holes closer until they become bound as a binary. As the orbit shrinks and becomes a hard binary, few-body scattering of nearby stars dominates. These stars have low angular momentum and are located in a volume of phase space called the ``loss cone" \citep{1996NewA....1...35Q,2002MNRAS.331..935Y,2003AIPC..686..201M}. 
If three-body scattering is efficient, the binary orbit shrinks by a factor of a few hundred, resulting in a binary which will start to emit gravitational waves and merge into one. SMBHB mergers are thought to be the most powerful gravitational wave sources in the Universe \citep{2003AnPhy.303..142H}. Of the three stages, the three-body scattering stage is widely thought to the longest, although some recent studies show the dynamical friction phase may now be the longest \citep{2015MNRAS.451.1868T, 2017ApJ...840...31D}; if the galaxy remnant is gas-free, non-rotating, and spherical, this stage can be longer than a Hubble time.  Here, the binary stalls without reaching the gravitational wave phase to finish the coalescence. This is the infamous {\it final parsec problem} \citep{2003AIPC..686..201M}. 

Although lingering questions remain for the dynamics of SMBHB, the consensus is that there are a number of ways to drive the SMBHB through the three-body scattering stage relatively quickly. One way is shape: triaxial and axisymmetric galaxy models keep the binary supplied with stars on centrophilic orbits to help the black hole binary merge \citep{2006ApJ...642L..21B,2006astro.ph..1520H, 2011ApJ...732...89K,2011ApJ...732L..26P,2013ApJ...773..100K,2014ApJ...785..163V}. Rotation also helps shrink the binary orbit \citep{2015ApJ...810..139H}, an effect often excluded from simulations despite the near-ubiquity of bulk rotation in classical and pseudobulges \citep{2004ARA&A..42..603K, 2012A&AT...27..221G,1992ApJ...399..462B,2009MNRAS.397..506K,2011MNRAS.416.1654B,2015ApJ...802L...3T}. A third intruder black hole from a subsequent galaxy merger can accelerate the coalescence of the SMBHB through the combined action of Kozai--Lidov resonances and gravitational wave emission \citep{2007MNRAS.377..957H,2016MNRAS.461.4419B,2017arXiv170906501R}. Finally, the viscous drag from a gaseous disk around the SMBHB may increase the binary's orbital decay rate \citep{2009ApJ...700.1952H,2009MNRAS.398.1392L}.

While it's true that centrophilic orbits in non-spherical and/or rotating systems are thought to be key in shrinking the SMBHB orbit, an analysis of the orbits that {\it are} scattered by the binary has not been undertaken. Instead, most studies analyze the orbits within a black hole-embedded primary galaxy before the second SMBH enters, keeping track of those with the potential of interacting with a SMBHB, if one were there \citep{2001ApJ...549..862H,2002ApJ...567..817H,2005MNRAS.360.1185J,2010ApJ...723..818H,2012MNRAS.422.1863B,2012MNRAS.419.1951V,2014MNRAS.445.1065R}. However, the perturbation from a second black hole may be significant enough to obliterate stable centrophilic orbital families as well as to generate new regions of stability in phase space.
It is important to understand the orbits that really do interact with the binary, as these orbits can influence the eccentricity and plane of the binary itself.


Here we examine the orbits of particles scattered by a SMBHB in an axisymmetric galaxy and compare this to the number and type of orbits predicted to interact with the binary within the initial conditions of a single SMBH-embedded galaxy model.

This paper is organized as follows. In Section 2 we describe the technique to identify the orbits which are predicted to interact with the binary versus the orbits which actually do interact. Our results are featured in Section 3, and the implications are covered in Section 4.

\section{Method}

Our base galaxy model, with 1 million particles, is in equilibrium, non-rotating, and flattened, having an axis ratio $c/a=0.75$ and a single SMBH. We introduce a second equal-mass SMBH in orbit within the base galaxy which inspirals and eventually merges with the central SMBH; we output 53 snapshots of the inspiral process, each snapshot containing the 6-d phase space information of each particle and SMBH in the system. Note that the orbital structure of the base model has previously been analyzed in  \citet{2015ApJ...811...25L}, and the inspiral dynamics was explored in  \citet{2013ApJ...773..100K}. Here, we track those particles that strongly interact with the SMBHB during the inspiral and analyze their orbits. For more detail about the orbital analysis technique or method of simulating the binary black hole inspiral and coalescence, please see \citet{2013ApJ...773..100K} and \citet{2015ApJ...811...25L}. In these models,
 we can scale the central black hole to the mass of that in the Milky Way, which sets the length unit to 20 pc, the velocity to $\sim 450$km/s, and time to $\sim 4\times 10^{4}$ years. In system units, the black holes have equal mass of 0.005, and each stellar particle has mass of $1\times 10^{-6}$.

We identify those particles that actually interact with the SMBHB by looking for a 
large energy boost in a particle in the simulation once the SMBHs are bound as a binary at $t=8$. We select the snapshots evenly from $t=0$ to 52, which are $t=0, 13, 26, 39$. Although at $t=0$ the binary is still soft, we analyze it as a comparison. 



\subsection{Particles with promise: Identifying centrophilic orbits in the base potential. }


As in \citet{2015ApJ...811...25L}, for $t=0, 13, 26, 39$, we freeze the potential of the base model at each snapshot, and run each particle to $t=52$. 
For each orbit, we record its
minimum separation from the SMBH, $r_\mathrm{min}$; the orbits with $r_{\rm min}$ less the hardening radius of the SMBHB (if it were present in the base model). The hardening radius is expressed as: $a_{h}=G\mu/4\sigma ^2$ 
\citep{1996NewA....1...35Q}. We select as model units the gravitational constant $G=1$, the reduced mass of the SMBHB $\mu=0.0025$ (since each black hole has a mass of 0.005), and the stellar velocity dispersion within the radius of influence of the SMBH $\sigma=0.4$. Therefore, here $a_{h}=0.004$.


\subsection{Scattered particles: tracking particles that interact with the SMBHB}

Before the SMBHB hardens, a star may gain or lose energy after interacting with the black holes. Since we are interested in interactions in general, we track all particles with a significant change in energy, and we define significant as greater than a 10\% change.


However, in general, the particles gaining energy are more than that losing energy, therefore the net energy all the interacting particles gain is positive, which should be equal to the energy the SMBHB loses. The total energy of the whole system is conserved to $\sim 10^{-4}$. For the snapshot of $t=0$, among the 1 million particles, $\sim$ 640000 particles gain energy and $\sim$ 360000 particles lose energy. However, for most particles the energy change is just due to the background galaxy potential evolving with time (see Figure \ref{density} in Section 3). Only the energy changing most particles are the ones that really interact with the SMBHB. Therefore we order the stellar particles according to the absolute value of their energy change percentage $\left|\Delta E/E\right|$ from high to low. The criterion we use here to find the ``scattered particles" with the SMBHB is that the total energy that the first $N_{r}$ energy-changing-most particles get equal to that the SMBHB loses. With this criterion, we get $N_{r}=14226$ for the snapshot of $t=0$. We note that among these 14226 particles only two hundred particles lose energy, all the other particles gain energy. In order not to miss some particles with high $\left|\Delta E\right|$ but low $\left|\Delta E/E\right|$. We also order the particles by $\left|\Delta E\right|$ from high to low and make use of the same criterion obtaining $N_{r}=11839$. The union of these two groups of particles are considered as the ``scattered particles". The total number is $\sim 14900$, mainly made of the ones obtained using $\left|\Delta E\right|$. For all the ``scattered particles", the $\Delta E/E$ is significant, which is between $10\%$ and $1000\%$, with minimum value $10\%$ and maximum value $41200\%$. We use the same method to obtain the ``scattered particles'' for other snapshots.

Figure \ref{number} shows the number of scattered particles, particles with promise, and common particles of the two groups at each snapshot of $t=0, 13, 26, 39$. The trend is particles with promise at any snapshot are more than that of scattered particles of the same snapshot, indicating that the binary does not scatter as many particles as it can -- it only scatter as many as it needs. It is natural that at earlier time it can and needs to scatter more. For example, from $t=0$ to $t=52$ the binary can scatter particles 1.5 times its mass, while from $t=39$ to $t=52$ it can only scatter half its mass. The common particles of the two groups are rare, usually less than 30\% of particles with promise of the same snapshot. It is inferred that both the particles with promise and scattered particles may be drawn from a larger collection, because even if they do not have many common particles, they still have the same characteristic quantities such as energy, $L_{z}$, etc., as shown in Section 3.

After identifying particles with promise and scattered particles for the four snapshots, we run the eight groups of particles for 100 dynamical times to get dominant frequencies along the principle axes for each particle. For the first 50 dynamical times, we obtain $f_{x1}$, $f_{y1}$, $f_{z1}$; for the second 50 dynamical times we obtain $f_{x2}$, $f_{y2}$, $f_{z2}$. We use Laskar's frequency mapping technique to classify orbits according to the dominant frequency ratios along the principle axes \citep{2010MNRAS.403..525V,2012MNRAS.419.1951V,2014ApJ...785..163V,2015ApJ...810...49V,2001ApJ...549..862H,2002ApJ...567..817H,2006astro.ph..1520H,2015ApJ...810..139H} and determine the orbital families present by integrating all particles to complete the phase coverage of the orbit. To identify chaotic orbits, we use the criterion that any two of fx, fy, fz must satisfy both $|f_{1}-f_{2}|>2f_{bin}$ and $|f_{1}-f_{2}|/f_{1}>10^{-1.22}$ \citep{2010MNRAS.403..525V}. As shown in Figure \ref{fft}, the main resonant orbital type is $fx:fy=1:1$. The way we identify this tube orbit is excluding all the chaotic particles, if a particle satisfies $fx/fy>0.95$ and $fx/fy<1.05$, it is considered as a $fx:fy=1:1$ orbit. The remaining orbits are resonant orbits. 
\begin{table*}[]
\caption{Energy table}
  \centering
  \begin{tabular}{c c c c c c c c}
\hline
 & Stellar Particles $E_{pot\_in}$ & Stellar Particles $E_{pot\_ext}$ & Stellar Particles $E_k$ & Stellar Particles $E_{tot}$ & BHB $E_{pot}$ & BHB $E_k$ & BHB $E_{tot}$\\ \hline
t=0 & -0.305 & -0.016 & 0.157 & -0.164 & 0.000 & 0.001 & 0.000   \\ \hline
t=52 & -0.290 & -0.017 & 0.159 & -0.146 & -0.034 & 0.017 & -0.017 \\ \hline
$\Delta$E & 0.016 & -0.001 & 0.002 & 0.018 & -0.034 & 0.017 & -0.018 \\ \hline
\end{tabular}
\label{energytable}
\end{table*}

\section{Results}

\begin{figure}
\centering
\includegraphics[width=75mm]{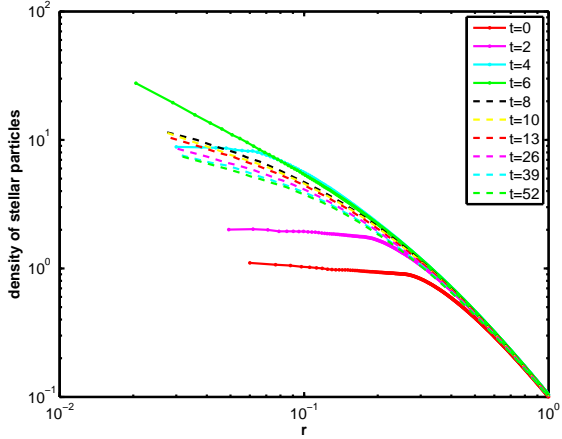}
\caption{Density of the central part of the two-SMBH model galaxy versus radius. This shows before the SMBHB becomes hard at $t=8$, the binary attracts stars nearby increasing its central density, and after the hardening at $t=8$ the interacting between the SMBHB and the stars decrease the density of the central part of the galaxy gradually. This also indicates that the region of influence of the SMBHB is $\sim 4$ times larger than that of one SMBH.}
\label{density}
\end{figure}

Table 1 lists the energy components and energy change of the stellar particles and the SMBHB in the two-SMBH model at $t=0$ and $t=52$. The first four columns show the stellar particles' potential relative to the stellar background $E_{pot\_in}$, their potential relative to the SMBHB $E_{pot\_ext}$, their kinetic energy $E_k$, and their total energy $E_{tot}$, respectively. The last three columns show the two SMBH's potential relative to each other $E_{pot}$, their kinetic energy $E_k$, and their total energy $E_{tot}$. From this table, we can notice that: the energy the SMBHB loses equal to that the stellar particles gain as expected; the stellar particles' $E_{pot\_in}$ changes a lot, while their $E_{pot\_ext}$ and $E_k$ do not change too much; the energy in the system mainly is given from $E_{pot}$ (the SMBHB's potential relative to each other), half obtained by their kinetic energy $E_K$, half obtained by the stellar particles' $E_{pot\_in}$. This indicates the stellar particles' main energy change between $t=0$ and $t=52$ is due to the SMBHB changing the stellar structure by interacting with them. Therefore we plot Figure \ref{density}.

\begin{figure}
    \centering
    \includegraphics[width=75mm]{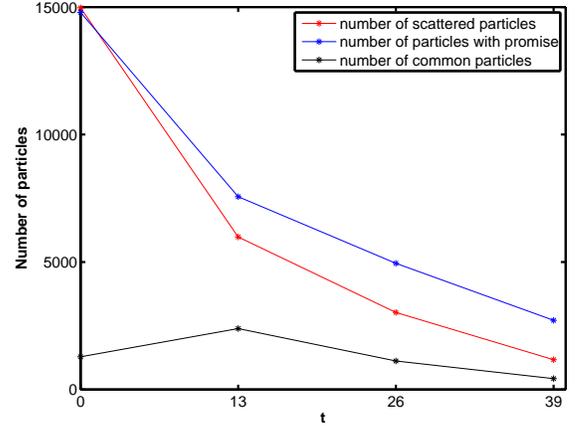}
    \caption{The number of scattered particles, particles with promise, and common particles of the two groups at each snapshot of $t=0, 13, 26, 39$. The trend is particles with promise at any snapshot are more than that of scattered particles of the same snapshot indicating that the binary does not scatter as many particles as it can -- it only scatter as many as it needs.}
    \label{number}
\end{figure}

\begin{figure}
\centering
\includegraphics[width=75mm]{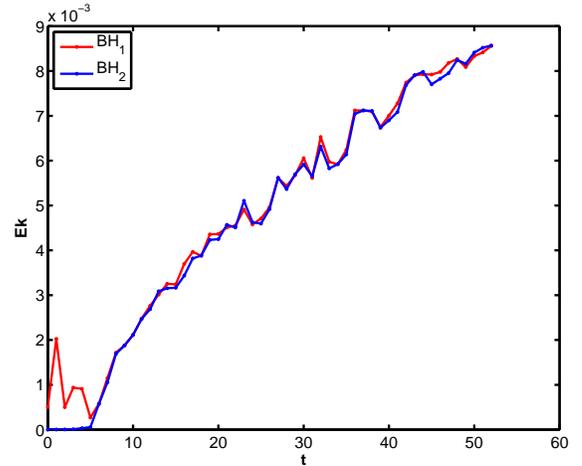}
\caption{Kinetic energy of the two SMBHs versus time. It shows that the kinetic energy of the two SMBHs increase with time. After $t=6$, the two SMBHs have nearly the same kinetic energy, which is as expected since they have equal masses.}
\label{Ekbh}
\end{figure}

\begin{figure*}[]
\centering
\includegraphics[width=50mm]{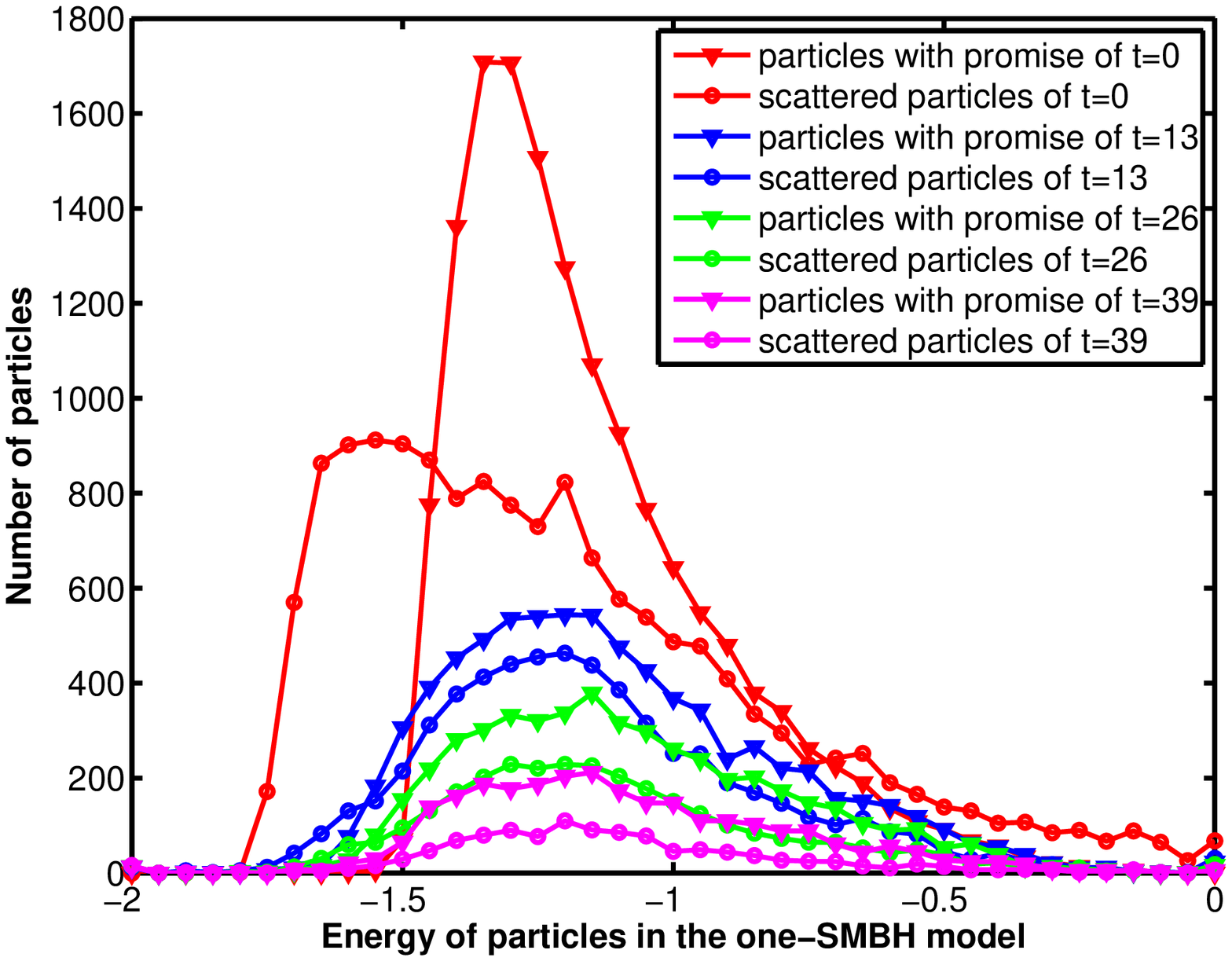}
\includegraphics[width=50mm]{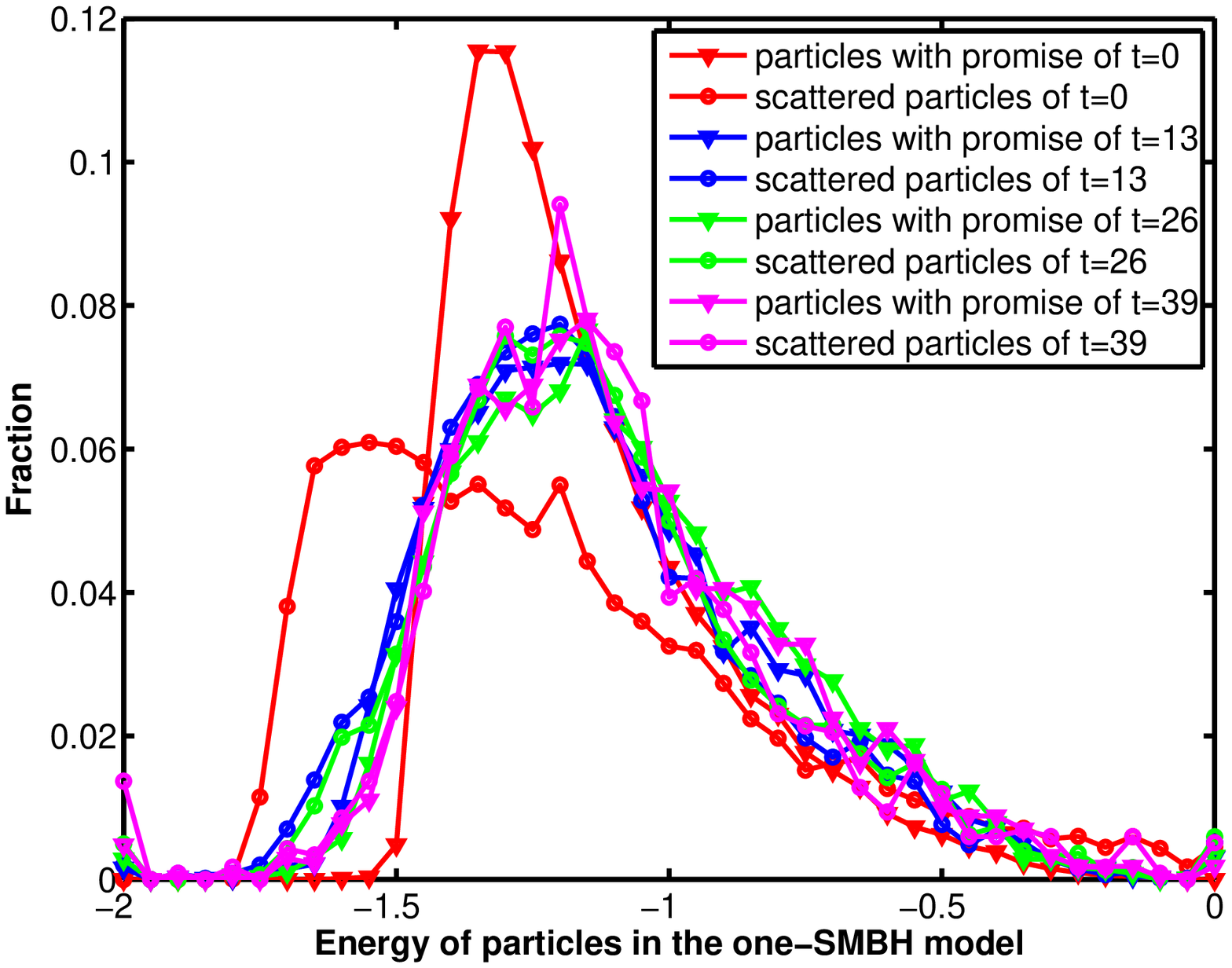}
\includegraphics[width=50mm]{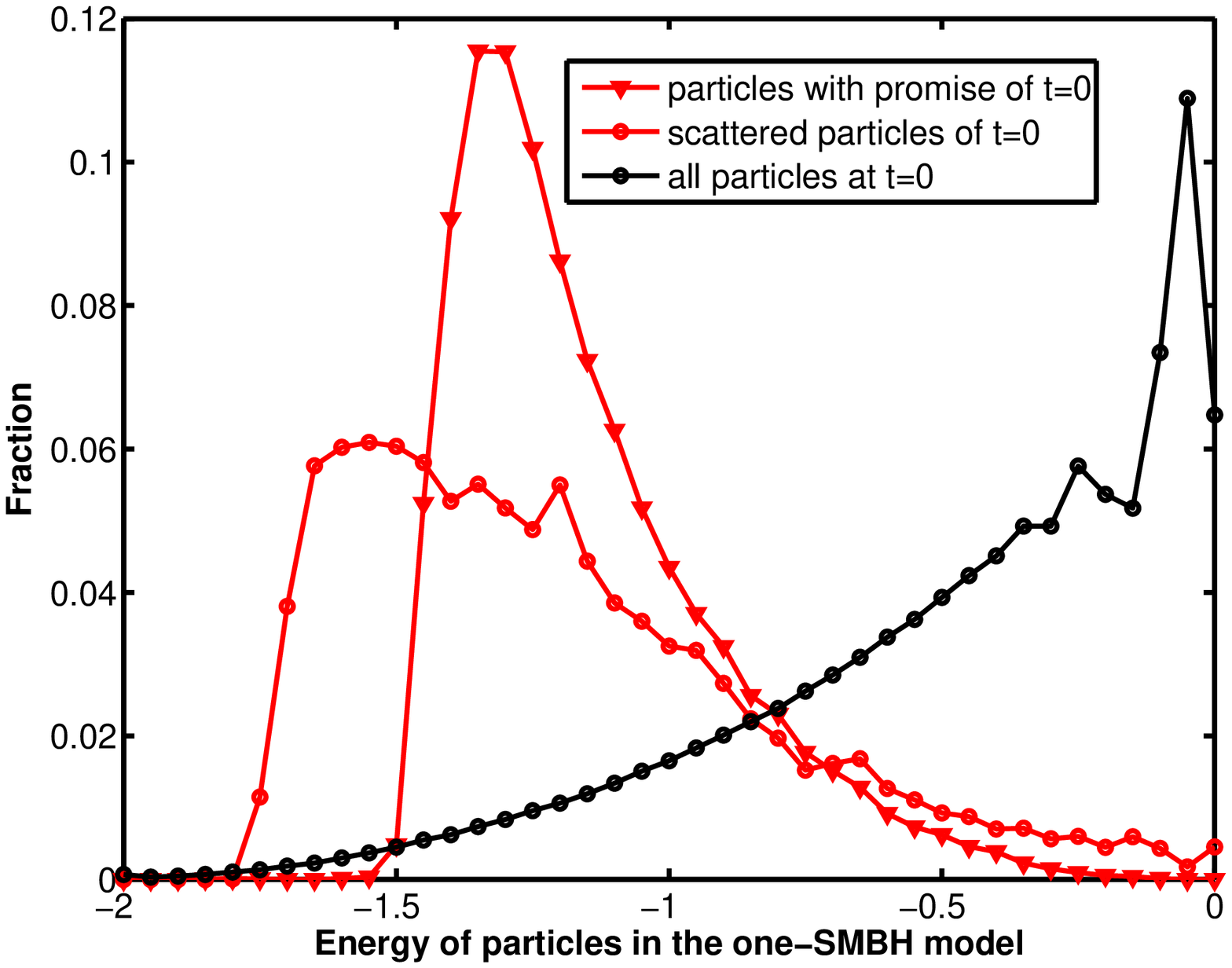}
\caption{Energy histogram. The left and middle panels show the energy distribution of particles with promise and scattered particles in the one-SMBH model for snapshots of $t=0, 13, 26, 39$, in number and fraction, respectively. The right panel of Figure \ref{Eplot} shows the energy distribution of the two groups and all the 1 million particles, in fraction, for $t=0$ snapshot. This figure shows in all snapshots except $t=0$ the particles from the two groups are from the same lower energy slice.}
\label{Eplot}
\end{figure*}

\begin{figure*}[]
\centering
\includegraphics[width=50mm]{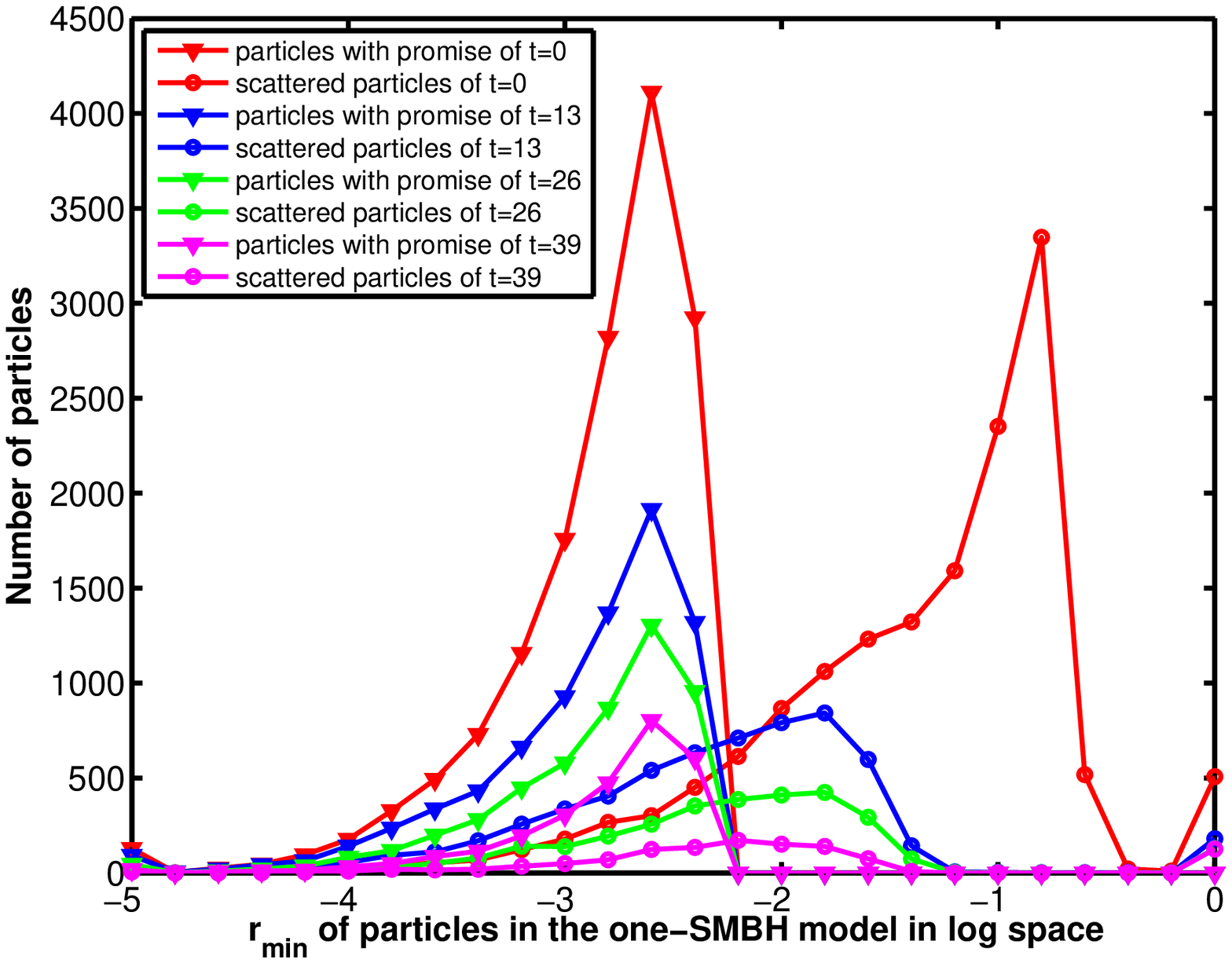}
\includegraphics[width=50mm]{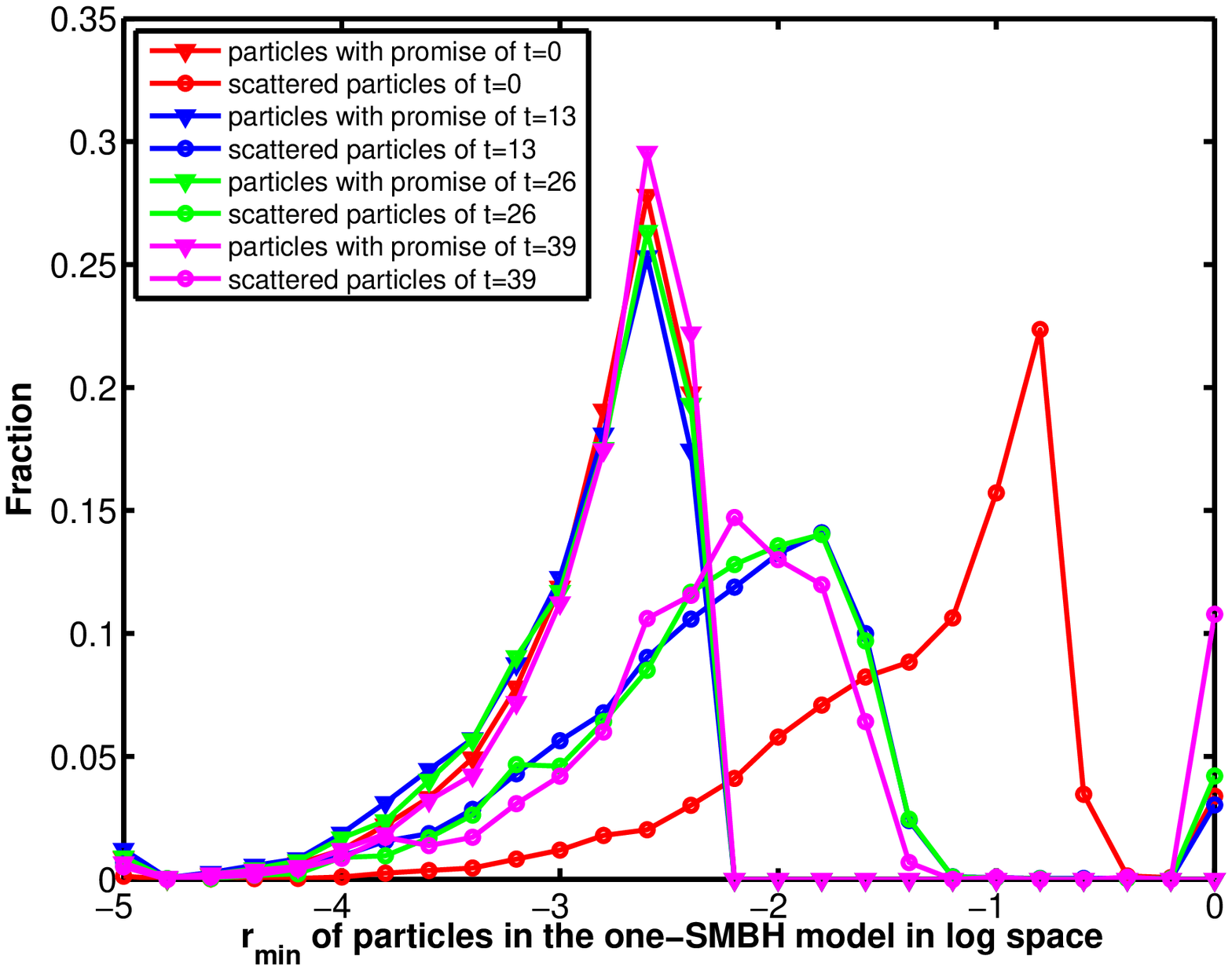}
\includegraphics[width=50mm]{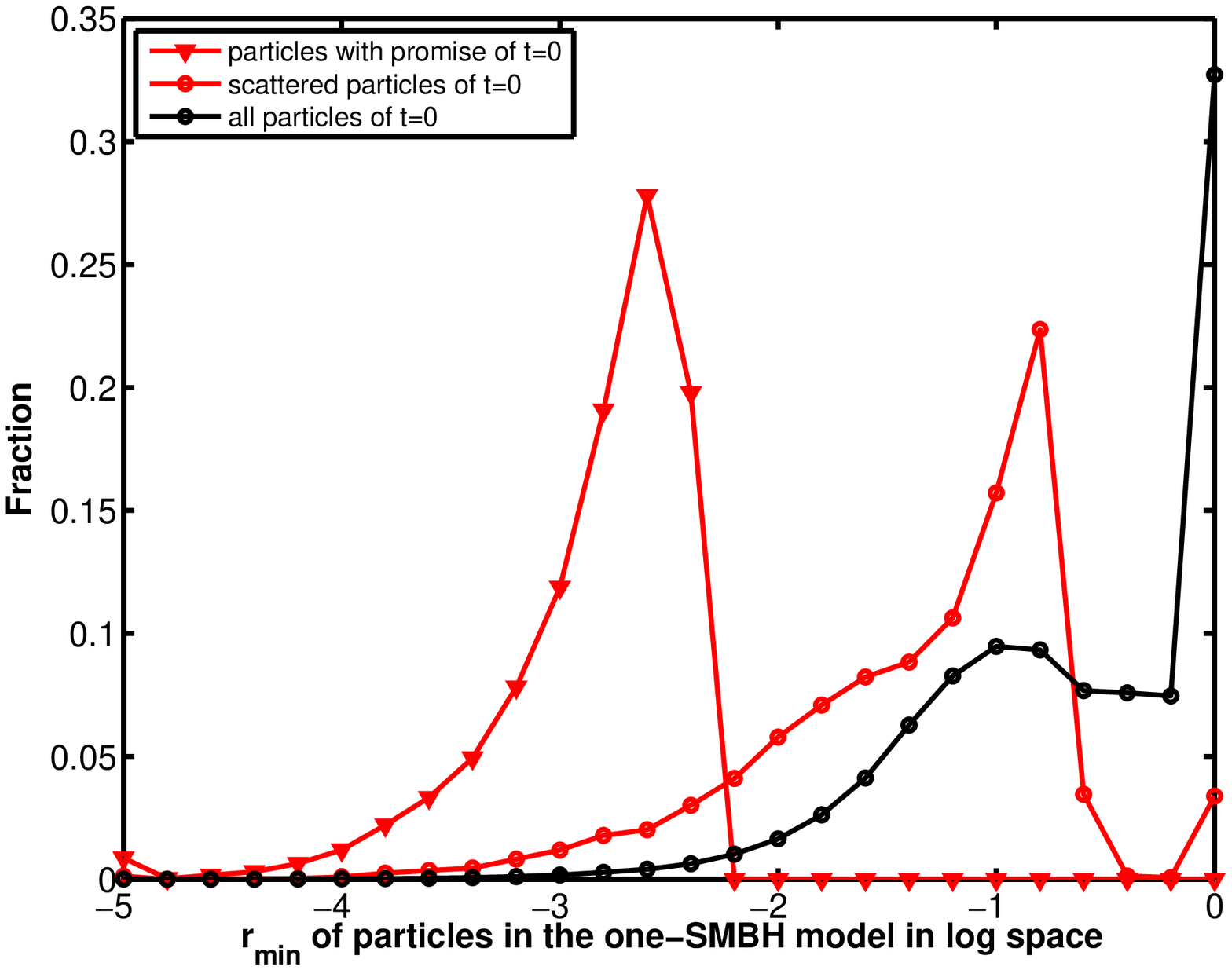}
\caption{Histogram for $r_\mathrm{min}$. The left and middle panels show the $r_\mathrm{min}$ histogram of particles with promise and scattered particles in the one-SMBH model for snapshots of $t=0, 13, 26, 39$, in number and fraction, respectively. The right panel shows the $r_\mathrm{min}$ histogram of particles with promise, scattered particles and all particles at $t=0$ in the one-SMBH model. This shows that both particles with promise and scattered particles have lower $r_\mathrm{min}$ than the group of all particles, and scattered particles have bigger  $r_\mathrm{min}$ than particles with promise.}
\label{rmin}
\end{figure*}

\begin{figure*}[]
\centering
\includegraphics[width=50mm]{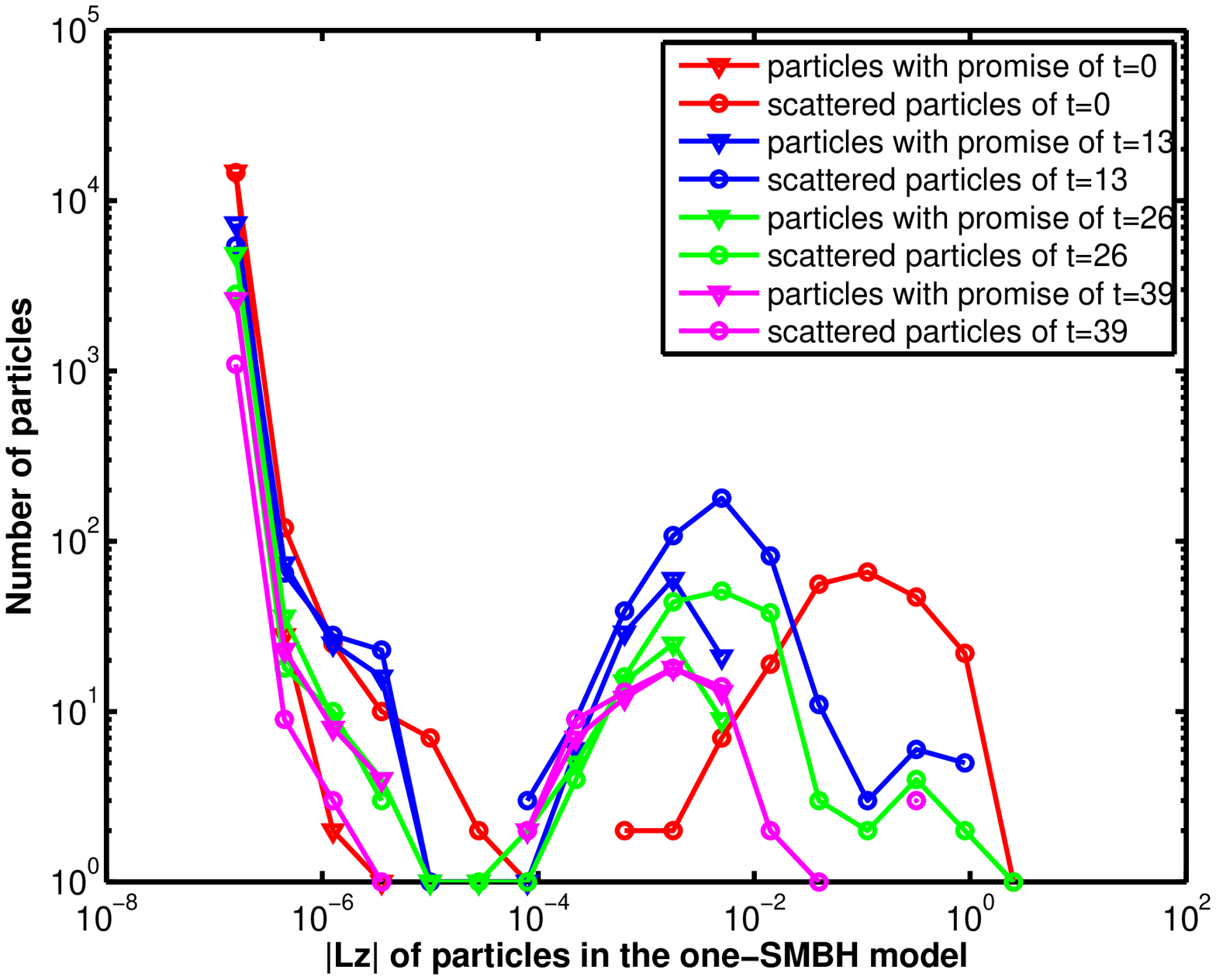}
\includegraphics[width=50mm]{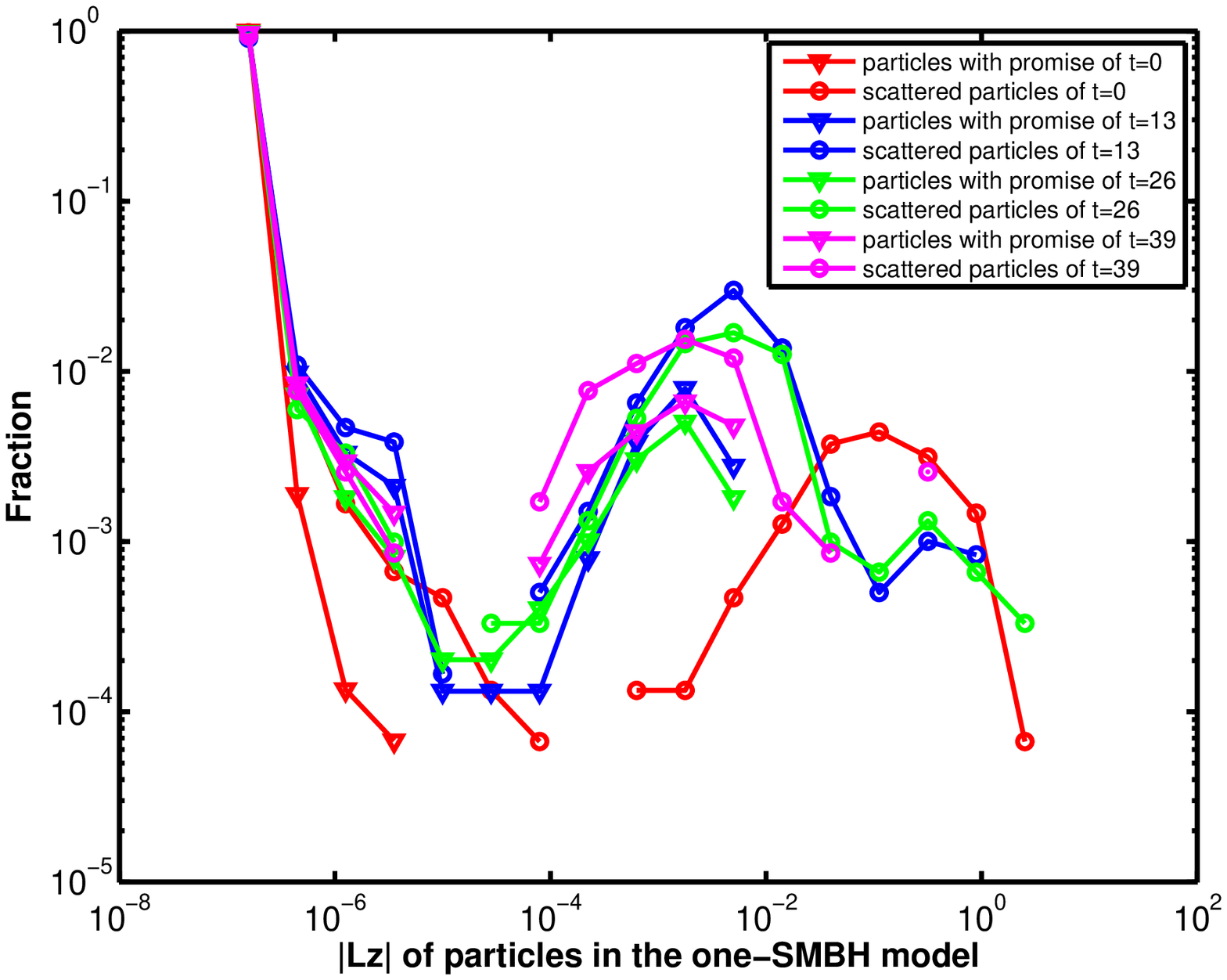}
\includegraphics[width=50mm]{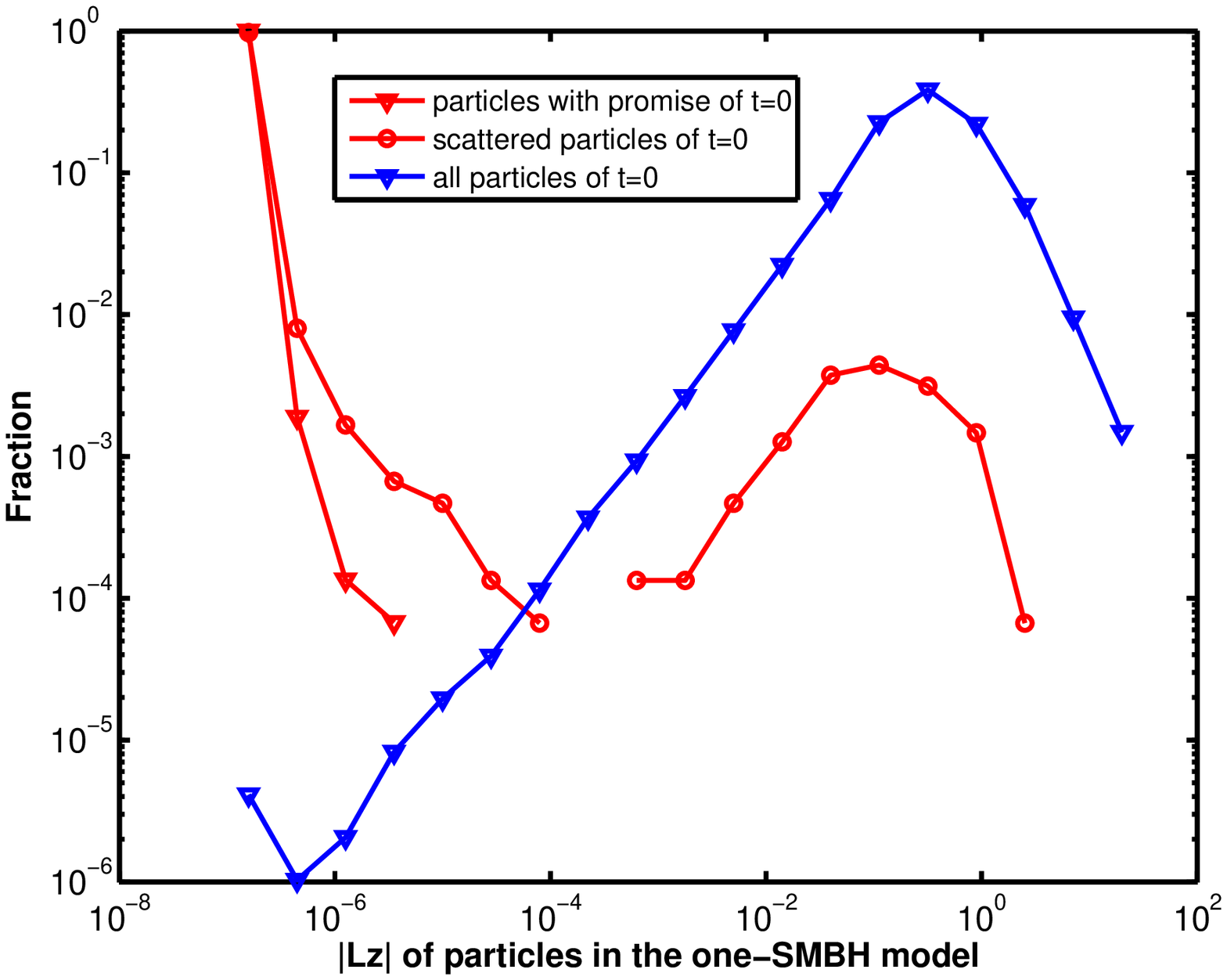}
\caption{Histogram for $L_{z}$. The left and middle panel of Figure \ref{lzmin} show the $L_{z}$ histogram of particles with promise and scattered particles in the one-SMBH model for snapshots of $t=0, 13, 26, 39$, in number and fraction, respectively. The right panel shows the $L_{z}$ histogram of particles with promise, scattered particles and all particles at $t=0$ in the one-SMBH model. This figure illustrates that both particles with promise and scattered particles have more than 90\% particles with $L_{z}$ around $10^{-7}$ much smaller than the group of all particles, which has $L_{z}$ peaking at around 1.}
\label{lzmin}
\end{figure*}

\begin{figure*}[]
\centering
\includegraphics[width=40mm]{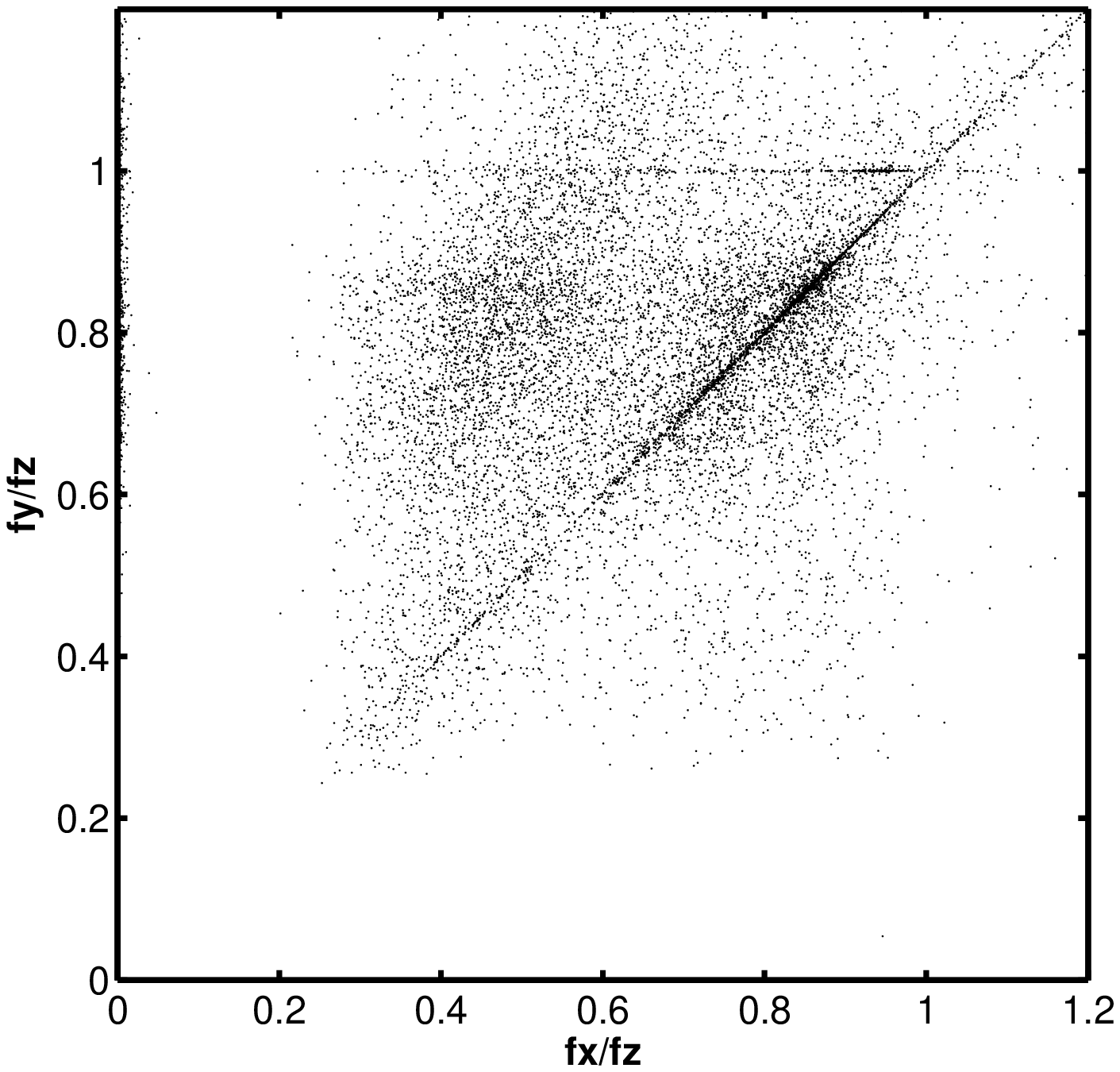}
\includegraphics[width=40mm]{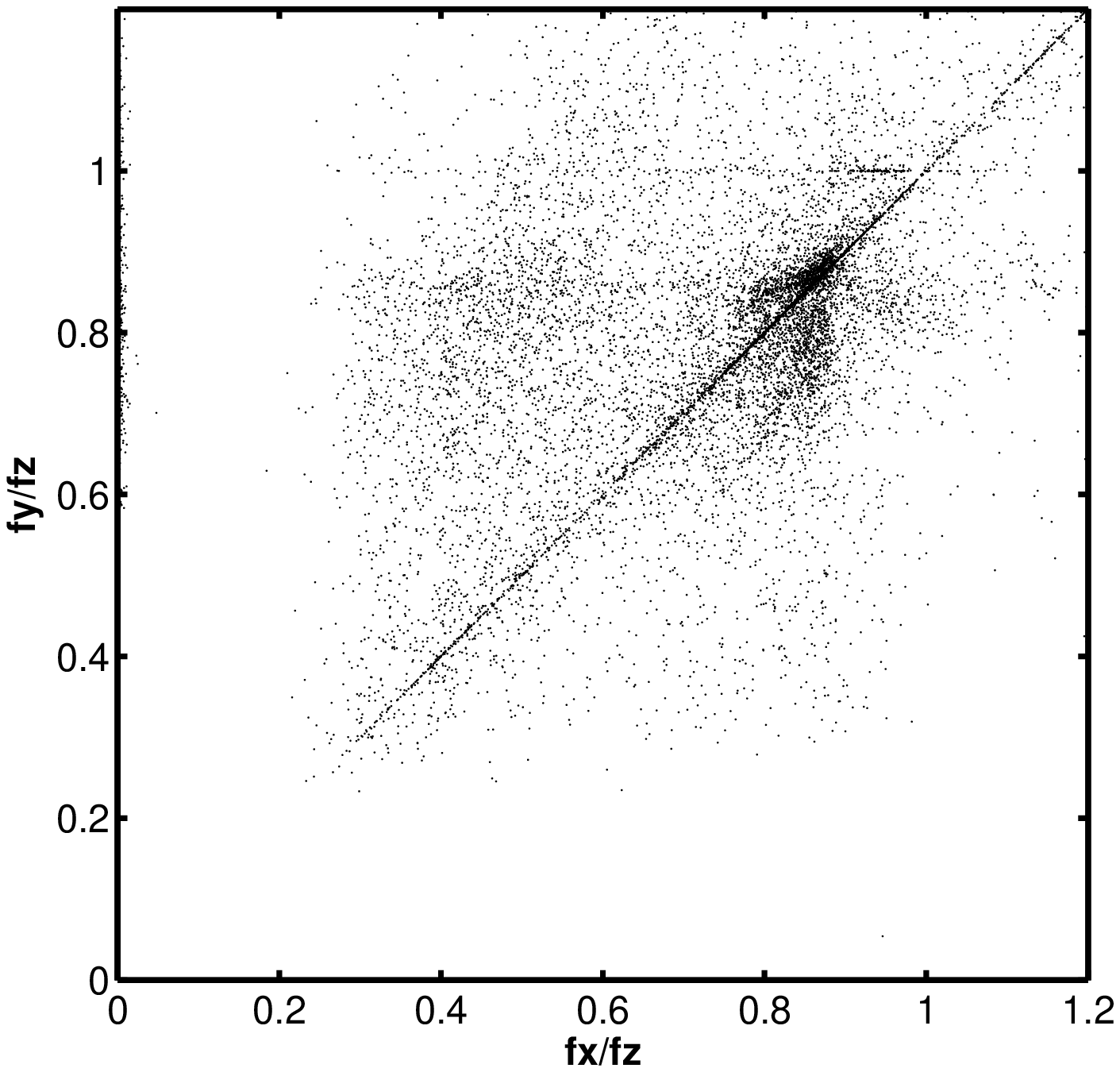}
\includegraphics[width=40mm]{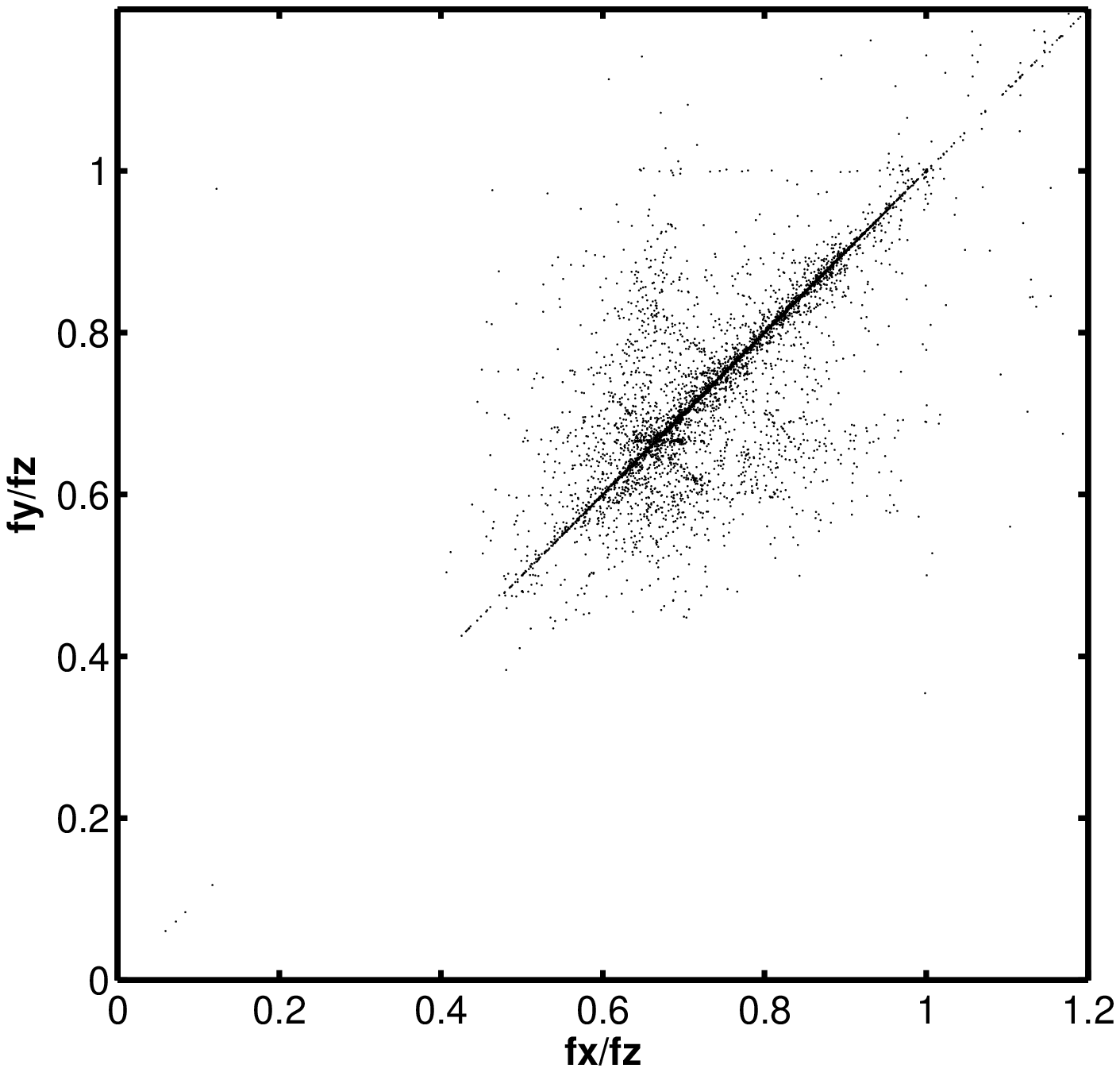}
\includegraphics[width=40mm]{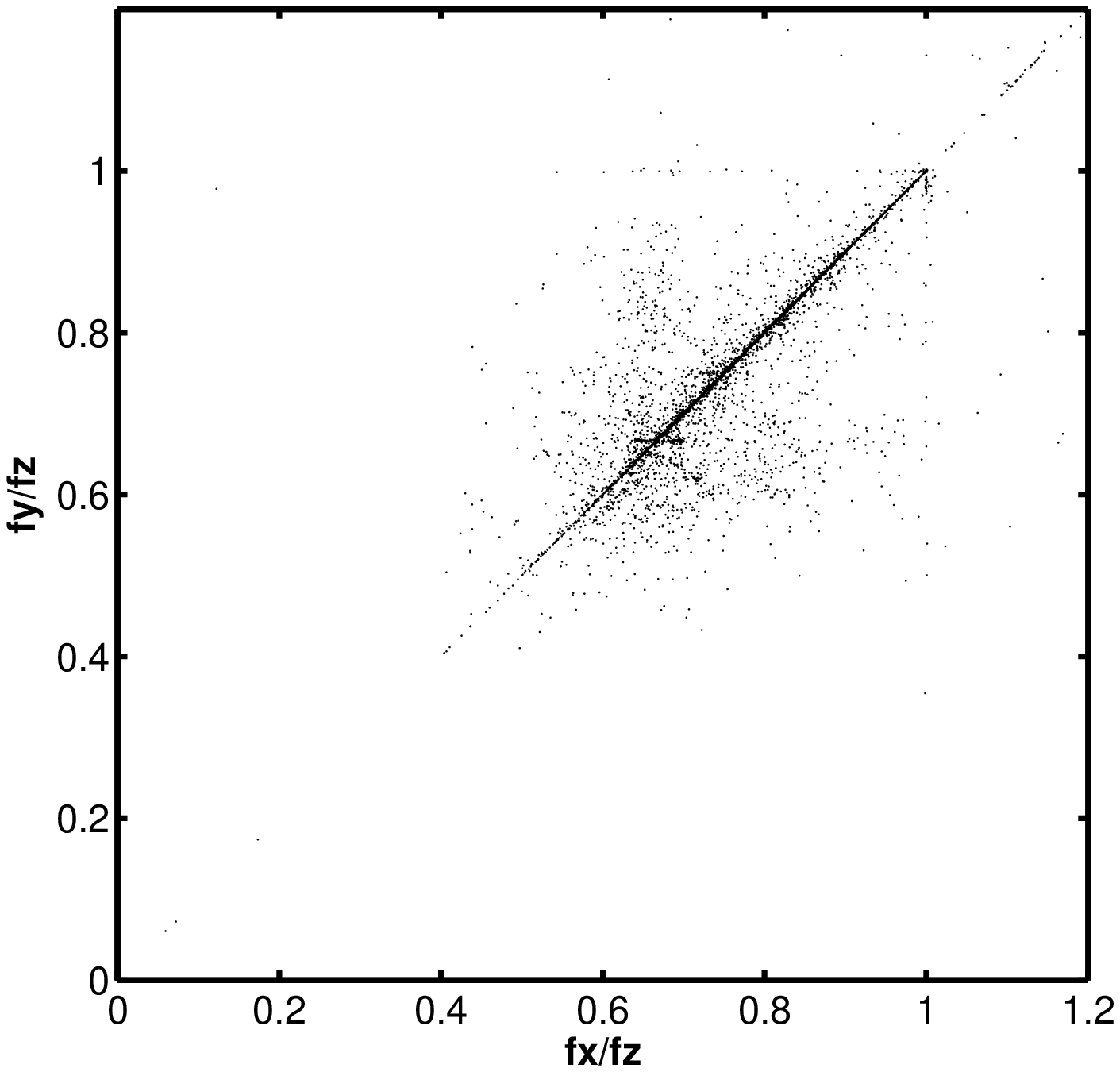}
\includegraphics[width=40mm]{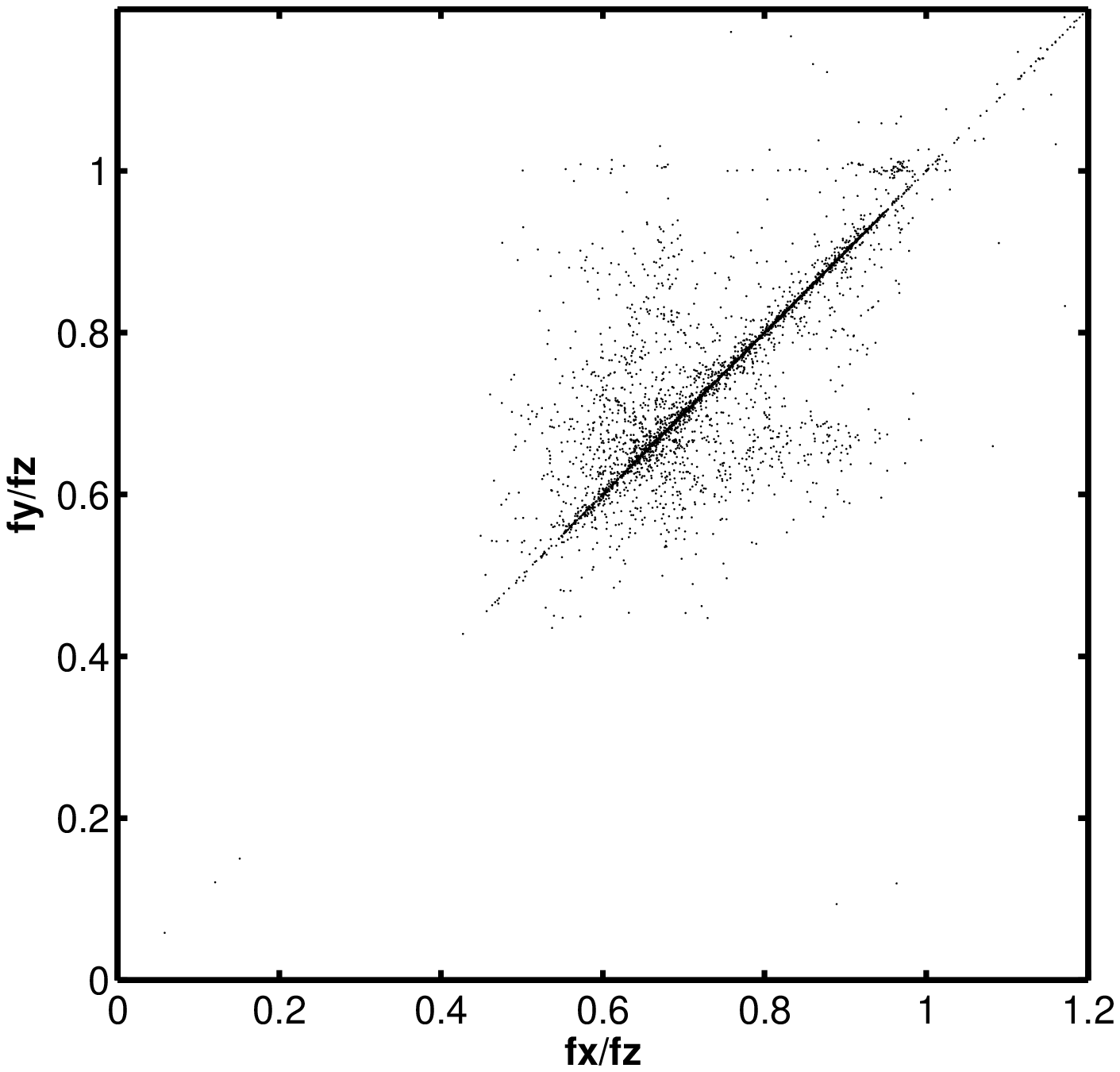}
\includegraphics[width=40mm]{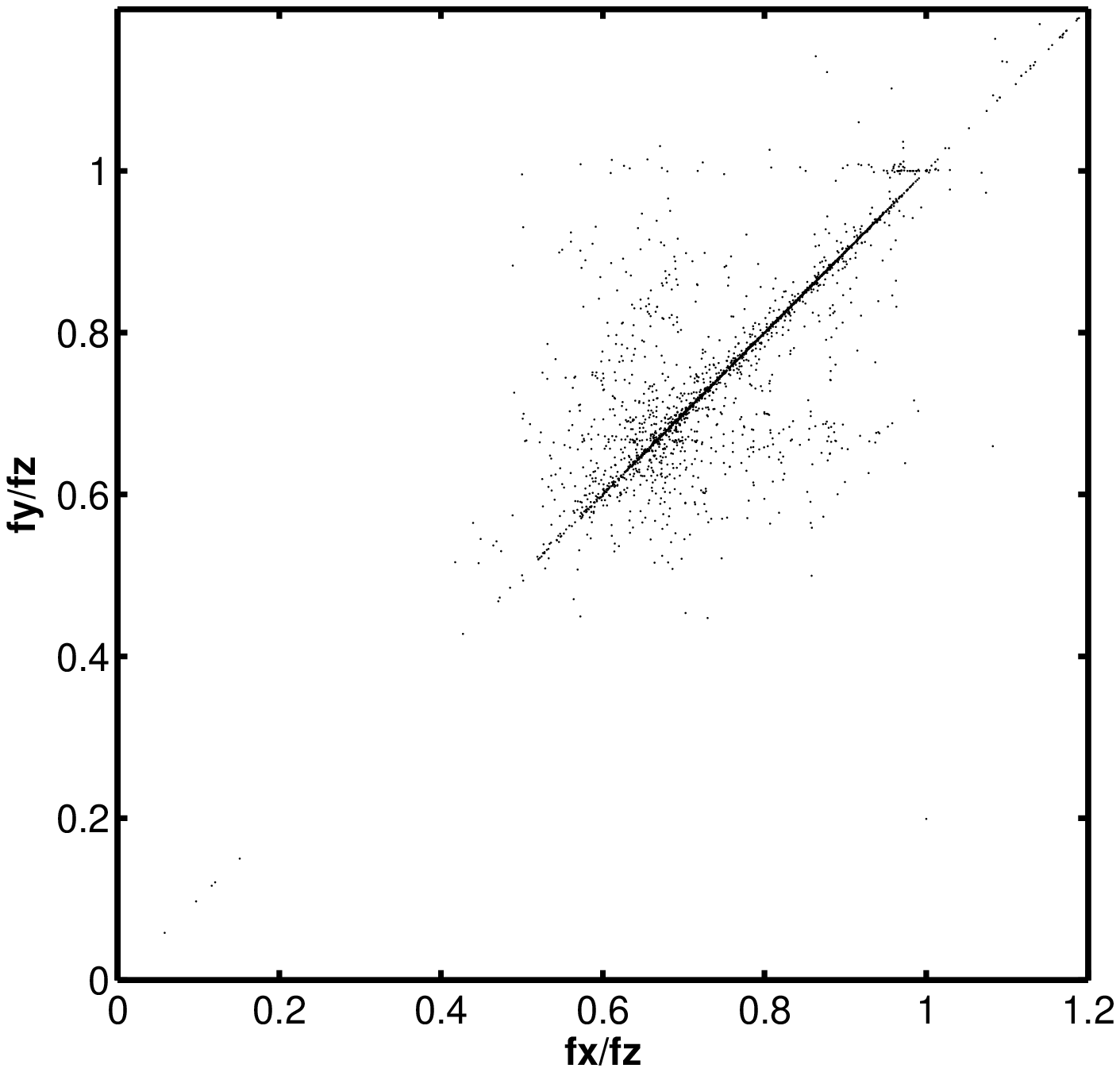}
\includegraphics[width=40mm]{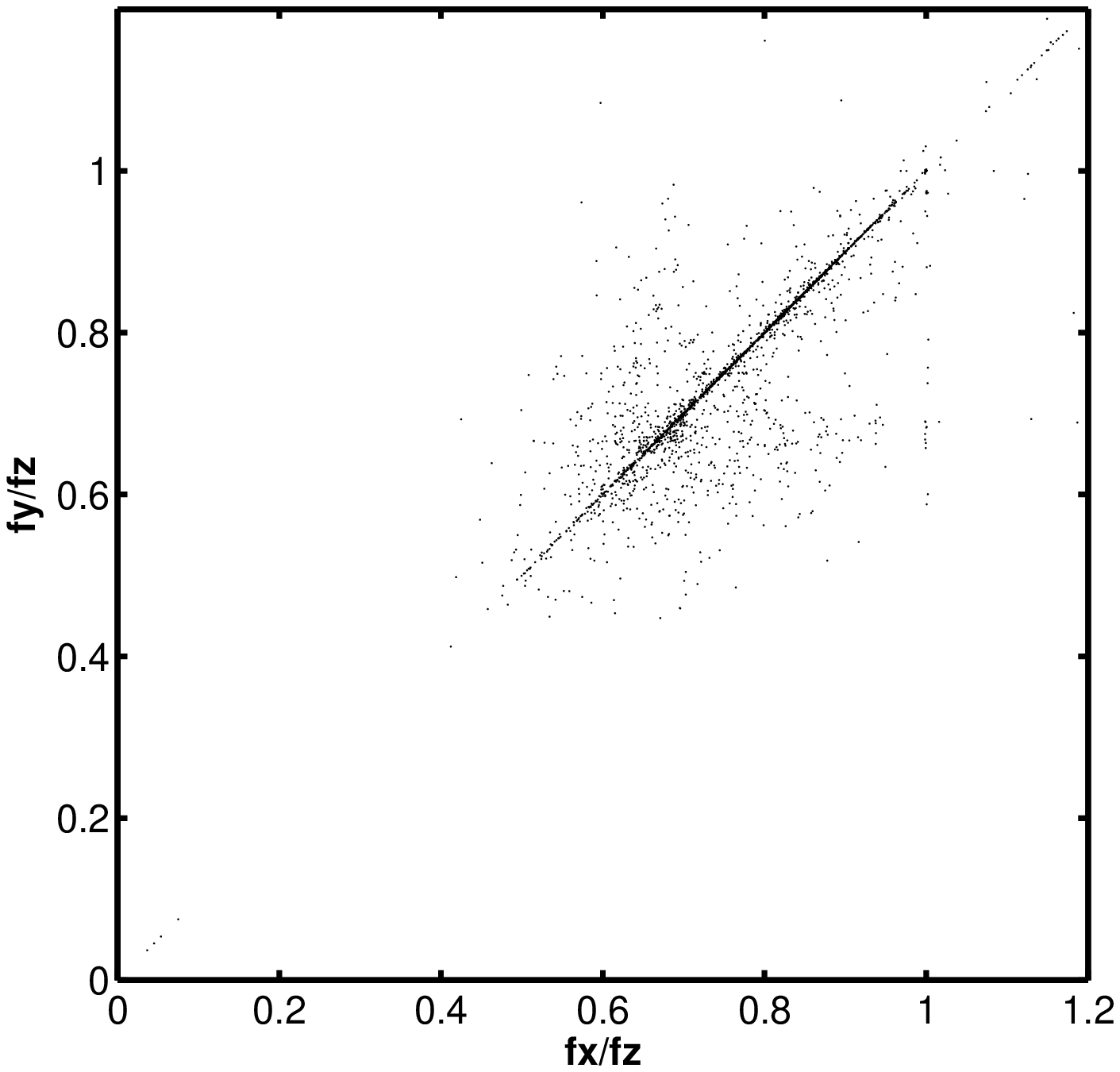}
\includegraphics[width=40mm]{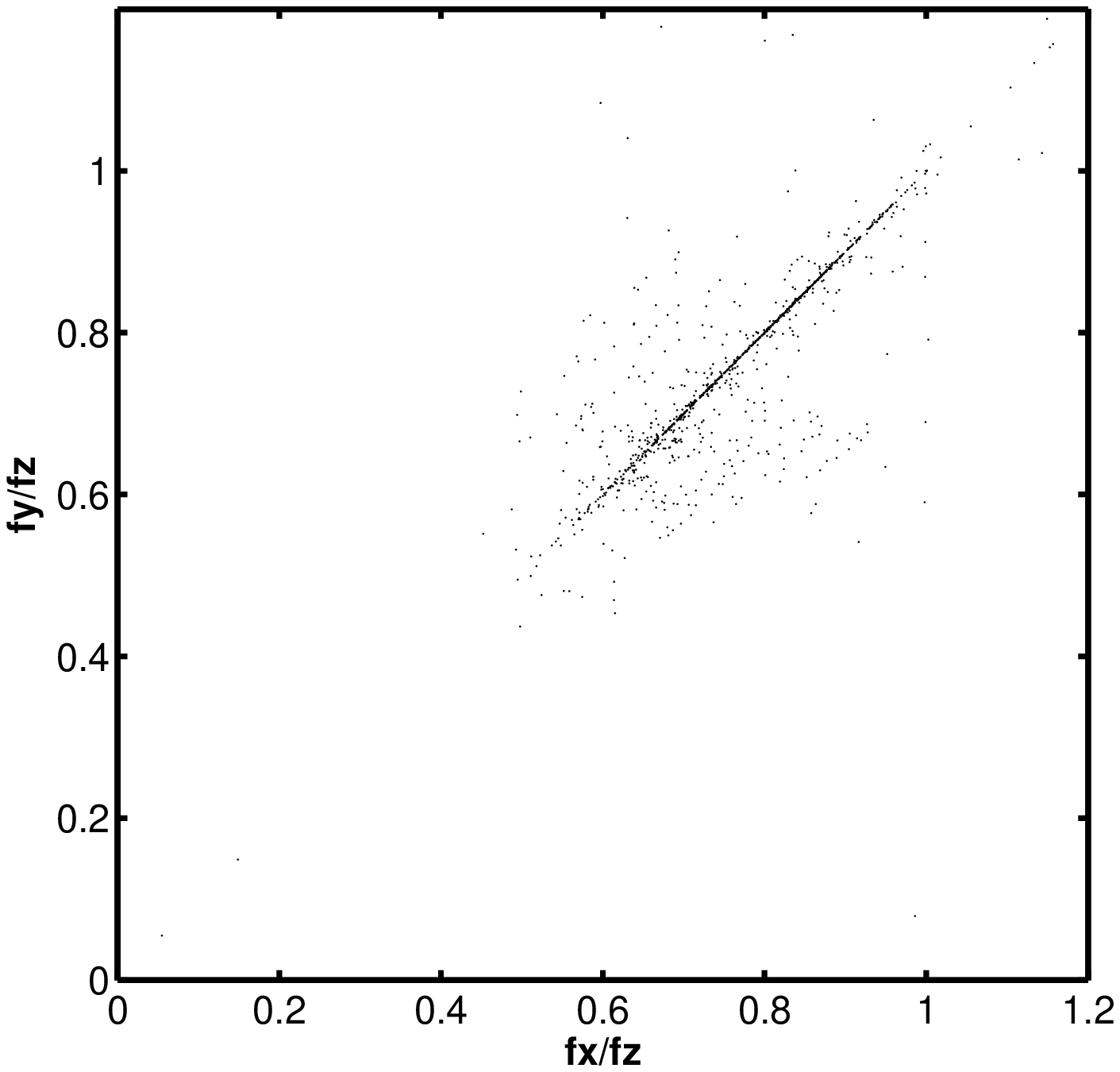}
\caption{Frequency maps. The upper panel shows the frequency maps of particles with promise of $t=0$, scattered particles of $t=0$, particles with promise of $t=13$, scattered particles of $t=13$, respectively. The lower panel shows the frequency maps of particles with promise of $t=26$, scattered particles of $t=26$, particles with promise of $t=39$, scattered particles of $t=39$, respectively.}
\label{fft}
\end{figure*}

\begin{figure*}[]
\centering
\includegraphics[width=73mm]{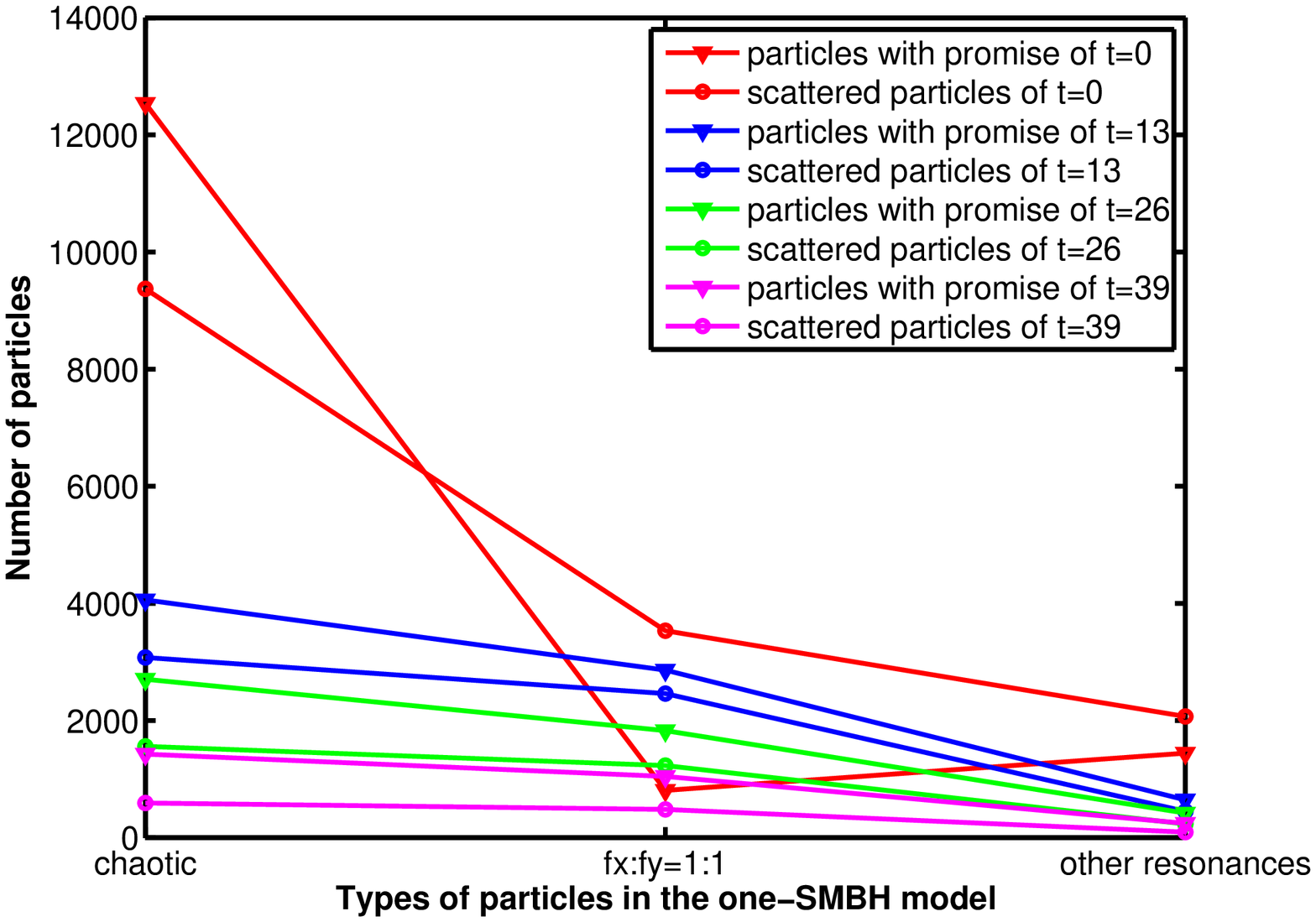}
\includegraphics[width=70mm]{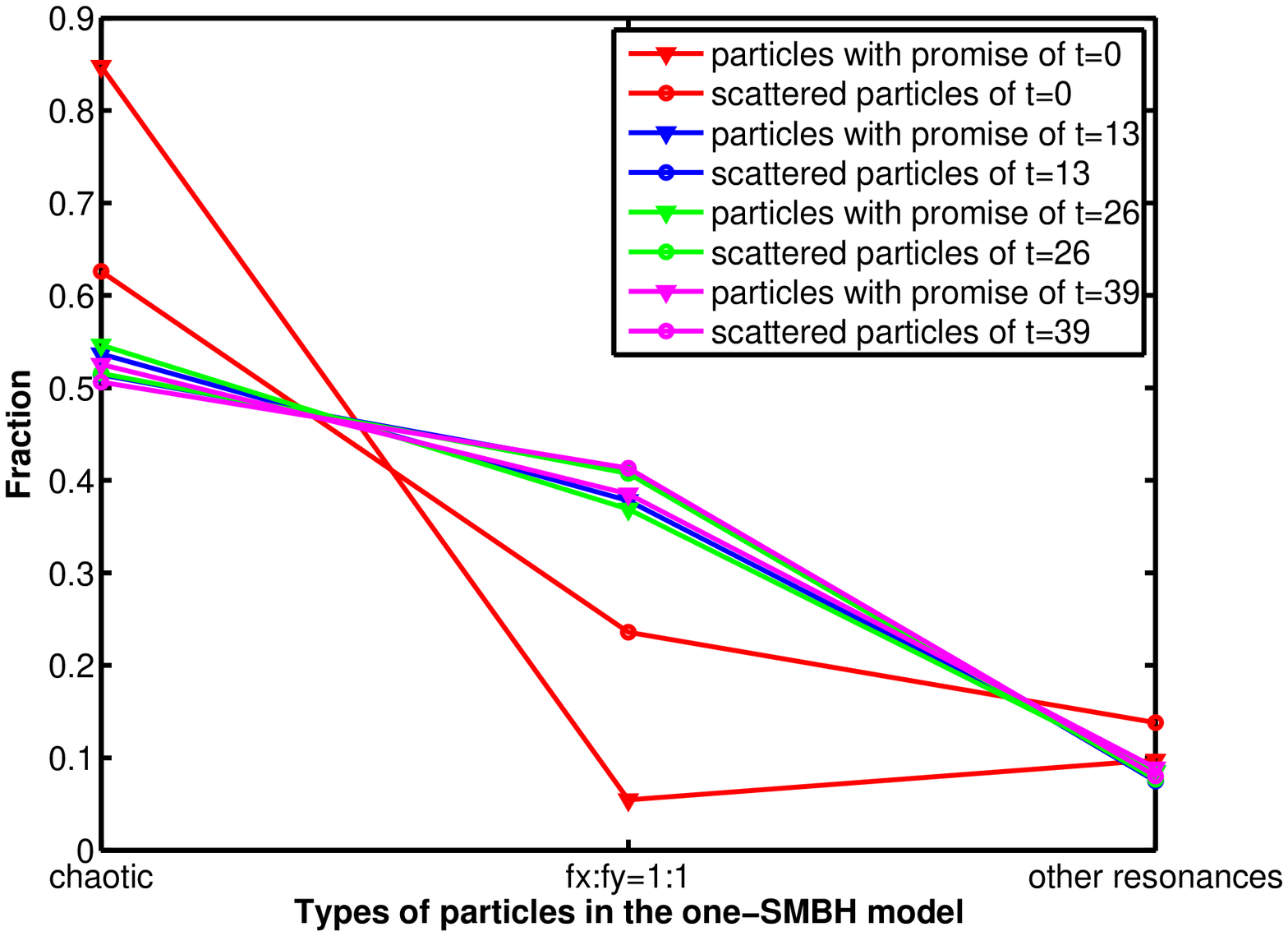}
\caption{Orbital classification. The left panel shows the number of particles in each type for the eight groups, the right panel shows the fraction of types in each group. }
\label{orbittype}
\end{figure*}

Figure \ref{density} shows the stellar density at snapshots of $t=0, 2, 4, 6, 8, 10, 13, 26, 39, 52$, respectively. The origin is set up to the center of mass of the SMBHB. The solid lines show the density before $t=8$, while dash lines show the ones at $t=8$ and after. It is seen that the density increases so much from $t=0$ to 6, then it decreases from $t=8$ to $t=52$ all along, though not too much. As we mentioned before, at $t=8$ the SMBHB becomes a hard binary, therefore this density trend can be explained this way: when the  SMBHB is soft, it mainly attracts nearby stars making the central density dramatically increase more than 10 times; while when the binary becomes hard, it three-body scatters the stars passing-by making the density decrease gradually. If the SMBH scales to that of our Milky Way, the radius of influence is 0.05 in system units. We noticed that the interaction of the SMBHB and surrounding stars influences the stellar density within 0.2 system units (4 times the radius of influence of a single SMBH). That means the region that the SMBHB can impact is at least several ($\sim 4$) times larger than the influence radius of one SMBH.     

Figure \ref{number} shows the number of scattered particles, particles with promise, and common particles of the two groups at each snapshot of $t=0, 13, 26, 39$. The trend is particles with promise at any snapshot are more than that of scattered particles of the same snapshot, indicating that the binary does not scatter as many particles as it can -- it only scatter as many as it needs. Also the earlier snapshots have more particles of either group than that in later snapshots as expected. 

Figure \ref{Ekbh} shows the kinetic energy of the two SMBHs increases with time. After $t=6$, the two SMBHs have nearly the same kinetic energy. Since they have equal masses, therefore they have equal velocity after $t=6$.

The left and middle panels of Figure \ref{Eplot} show the energy histogram of particles with promise and scattered particles in the one-SMBH model for snapshots of $t=0, 13, 26, 39$, in number and fraction, respectively. The trend is for any energy bin in any snapshot except $t=0$ the number of particles of promise is bigger than that of scattered particles. The amazing thing is in the fraction panel all the lines except the ones for $t=0$ lie on each other perfectly, which shows in all snapshots except $t=0$ the particles from the two groups are from the same energy slice. The right panel of Figure \ref{Eplot} shows the energy distribution of the two groups and all the 1 million particles, in fraction, for $t=0$ snapshot. We notice that for the ``all particles" group the energy peaks at $E=-0.1$, which is much bigger than the peak position of particles with promise at $E=-1.3$ and the peak position of scattered particles at $E=-1.6$. This indicates that the particles with promise and scattered particles both belong to lower energy slice. Though not showing energy of all particles in other snapshots, particles with promise and scattered particles in other snapshots are also from lower energy slice.

The left and middle panels of Figure \ref{rmin} show the $r_\mathrm{min}$ histogram of particles with promise and scattered particles in the one-SMBH model for snapshots of $t=0, 13, 26, 39$, in number and fraction, respectively. In the number panel particles with promise have more particles in every $r_\mathrm{min}$ bin. In the fraction panel all the lines of particles with promise are perfectly overlying with each other peaking at $r_\mathrm{min}=10^{-2.8}$, also all the lines of scattered particles are nearly overlying with each other peaking at $r_\mathrm{min}=10^{-2.0}$ except the one of $t=0$, which indicates that scattered particles can have bigger $r_\mathrm{min}$. The right panel shows the $r_\mathrm{min}$ histogram of particles with promise, scattered particles and all particles at $t=0$ in the one-SMBH model. It is seen that the two groups have much smaller $r_\mathrm{min}$ than the group of all particles has.

$L_{z}$ is conserved in axisymmetric galaxies. The left and middle panel of Figure \ref{lzmin} show the $L_{z}$ histogram of particles with promise and scattered particles in the one-SMBH model for snapshots of $t=0, 13, 26, 39$, in number and fraction, respectively. We can see from both panels that for both groups more than 90\% particles have very small $L_{z}$ around $10^{-7}$. The right panel shows the $L_{z}$ histogram of particles with promise, scattered particles and all particles at $t=0$ in the one-SMBH model, illustrating that both groups have much lower $L_{z}$ than the group of all the particles does, which has $L_{z}$ peaking at around 1. 

Figure \ref{fft} shows the frequency map of the eight groups of particles. The upper panel shows the frequency maps of particles with promise of $t=0$, scattered particles of $t=0$, particles with promise of $t=13$, scattered particles of $t=13$, respectively. The lower panel shows the frequency maps of particles with promise of $t=26$, scattered particles of $t=26$, particles with promise of $t=39$, scattered particles of $t=39$, respectively. It is seen that the main resonant orbital type is $fx:fy=1:1$. We can also notice this in Figure \ref{orbittype}, the left panel of which shows the number of particles in each type for the eight groups, the right panel of which shows the fraction of types in each group. Except the scattered particles of $t=0$, all other groups have chaotic orbits for the most, then $fx:fy=1:1$ orbits, then other resonant orbits. It is also no surprising that for each type, the number of particles decreases from $t=13$ to $t=39$. While also except $t=0$, all other groups have the same fraction for the three orbital types -- all have chaotic orbits around 50\%, $fx:fy=1:1$ resonant orbits around 40\%, other resonant orbits around 10\%. This again verifies that the prediction groups are the same groups of particles with the scattered particles, except $t=0$ groups. Meanwhile, it shows that chaotic orbits are the most particles both in prediction and scattered groups, which is due to both groups very close to the SMBHB as expected.

\section{Conclusions and Discussion}

Several ways are provided to solve the `` final parsec problem", which is thought not a problem in a realistic galaxy, but the detail is not clear. We want to know if the particles which in a one-SMBH-embedded-in galaxy can get very close to the center(i.e. $r_\mathrm{min}<a_h$) still can do that in the same galaxy with two-SMBH-embedded-in making the SMBHB merge. Two equal mass SMBHs in axisymmetric galaxy model with 1 million particles and $c/a=0.75$ in \citep{2013ApJ...773..100K} can merge into one within 2.4 Gyr. Since the energy the SMBHB loses equal to that the stellar particles obtain, we define the energy changing most particles as the `` scattered particles" with the SMBHB. At the same time we fix this model's potential at the snapshots of $t=0, 13, 26, 39$, making only one SMBH at the center of the galaxy and predict the ``particles with promise" that can be potentially interacting with the SMBHB by checking the particles' $r_\mathrm{min}$ before $t=52$ as in \citet{2015ApJ...811...25L}. Then we rerun the two groups of particles for 100 dynamical times to obtain each particle's frequency and the z component of angular momentum $L_{z}$. We use Laskar's frequency mapping technique to classify orbits according to the dominant frequency ratios along the principle axes.

To summary, we find that after the SMBHB hardens the particles with promise and scattered particles are drawn from the same collection of phase space, although they do not have too many particles in common. Between $t=0$ to $t=52$ (the SMBHB hardens at $t=8$) the particles' energy change comes from the variation of the central density of the galaxy not the directly interaction with the SMBHB. Some main conclusions are as follows. 

1. The number of scattered particles in all snapshots except the one at t=0 is less than that of particles with promise, which maybe caused by $a_{h}$ decreasing after the SMBHB's hardening, while in the prediction we apply a constant $a_{h}$.

2. The energy and $L_{z}$ distribution of particles with promise and scattered particles except t=0 are nearly the same, showing that the two groups of particles are drawn from the same collection in phase space.

3. The stellar central density in the two SMBHs' model increases before the binary becomes hard and decreases after it becomes hard. It is suggested that when the SMBHB is soft, it mainly attracts nearby stars making the central density dramatically increase more than 10 times; while when the binary becomes hard, it three-body scatters the surrounding stars making the density decrease gradually.

4. The $r_\mathrm{min}$ of scattered particles is bigger than that of particles with promise. 

5. Except $t=0$, for both particles with promise and scattered particles, nearly 50\% of them have chaotic orbits, 40\% have fx:fy=1:1 orbits, the remaining 10\% have other resonant orbits.

6. Because of the larger mass, the SMBH model has a radius of influence $\sim 4$ times larger than in the single SMBH model, which allows the binary to draw from a larger reservoir of orbits to scatter. 

7. After hardening, the two SMBHs have the same kinetic energy.

\acknowledgements

B.L. acknowledges Sarah Bird for help editing the paper. B.L. also acknowledges support from the International Postdoctoral Exchange Fellowship provided by the Office of China Postdoctoral Council. The simulations were performed on the facilities of the Center for High Performance Computing at Shanghai Astronomical Observatory. F. K. acknowledges the dedicated GPU cluster ACCRE at the Advanced Computing Center for Research and Education at Vanderbilt University, Nashville, TN, USA, on which the initial simulations in \citet{2013ApJ...773..100K} were performed.

\bibliography{twobh}
\end{document}